# Direct observation of strong correlations near the band insulator regime of Bi misfit cobaltates

V. Brouet[1], A. Nicolaou[1], M. Zacchigna[2], A. Tejeda[3], L. Patthey[4], S. Hébert[5], W. Kobayashi[5], H. Muguerra[5] and D. Grebille[5]

[1]*Lab. Physique des Solides, Université Paris-Sud XI, UMR8502, Bât 510, 91405 Orsay (France)*
[2] *CNR - INFM, Lab. Nazionale TASC c/o Area Science Park, s.s. 14 Km. 163.5, I-34012 Basovizza (TS) (Italy)*
[3]*Lab. Matériaux et phénomènes quantiques, UMR 7162, CNRS, Université Paris Diderot, BP 7021, 75205 Paris (France)*
[4]*Swiss Light Source, Paul Scherrer Institut, CH-5234 Villigen (Switzerland)*
[5]*Laboratoire CRISMAT, UMR 6508 CNRS et Ensicaen, 14050 Caen (France)*

We present the first angle-resolved photoemission (ARPES) measurement of Fermi Surface in the "misfit" cobaltate $[Bi_2Ba_2O_4].[CoO_2]_{~2}$. This compound contains the same triangular Co planes as Na cobaltates, but in a different 3D environment. Our data establish the similarity of their electronic structure. We propose that the peculiar lineshape of all cobaltates is of the "peak-dip-hump" type, due to strong many-body effects. We detect a progressive transfer of spectral weight from the quasiparticle feature near $E_f$ to a broad hump in misfit phases where Ba is replaced by Sr or Ca. This indicates stronger many-body interactions in proximity of the band insulator regime, which we attribute to the presence of unusual magnetic excitations.

Describing the motion of holes on a triangular Co lattice is a fundamental challenge, because of the interplay between many complex phenomena. Strong correlations coexist with magnetic frustration, orbital degeneracy and possibly charge ordering effects. This situation occurs in Na and misfit cobaltates, where a triangular array of Co is embedded in edge-sharing oxygen octahedra, forming $CoO_2$ slabs. Paradoxically, experimental investigations in $Na_xCoO_2$ seem to indicate stronger correlations near the band insulator ($x=1$) than near the Mott insulator (expected at $x=0$, where there is one spin ½ hole per site). Indeed, Curie-Weiss (CW) susceptibilities and magnetic orders are found at $x>0.7$, whereas apparently simple metallic behaviors with Pauli susceptibilities are found at $x<0.5$ [1]. This unexpected contrast clearly calls for an explanation. It was sometimes assigned to the role of Na, whose potential may localize electrons on some Co sites in Na rich phases [2,3]. However, misfit cobaltates, which are Na free, also exhibit metallic phases with weak to strong CW behaviors [4-6] and even a progressive localization of carriers for small hole contents [5,6] not observed in Na phases [7]. In addition, both families exhibit high thermoelectric powers near the band insulator regime [4-8], with unusual magnetic field dependences [8].

ARPES is a sensitive probe of correlation effects, because removing one electron from a metal in the photoemission process reveals the way it was bounded to its environment. To clarify the nature of correlations through the phase diagram of cobaltates, we present an ARPES investigation of three misfit cobaltates, ranging from the metallic to the weakly hole-doped insulating regime. This study supports the idea that the ARPES lineshape of cobaltates cannot be simply understood in terms of band effects, as previously assumed [9], but that many-body effects are essential. We propose that the lineshape is of the "peak-dip-hump" (PDH) type, a structure often observed in correlated systems [10]. A coherent quasiparticle (QP) peak is followed by an incoherent tail, the "hump". The QP is an electron "dressed" by collective excitations and the hump corresponds to excited states related to this "dressing". We observe a transfer of spectral weight from the QP peak to the hump near the insulating regime. This indicates increasing correlation effects, which would be expected near a Mott insulator, but are more surprising here. As magnetic correlations are known to get stronger in this part of the phase diagram, this strongly suggests that they play an essential role in the "dressing" of the QP and gives indications to model these excitations.

In Bi misfit cobaltates, the $CoO_2$ slabs are stacked with 4 pseudoquadratic rock-salt (RS) layers, of the type $Ae$-O/Bi-O/Bi-O/$Ae$-O ($Ae$=Ba, Sr, Ca). Their ideal formula is $[Bi_2Ae_2O_4]^{RS}.[CoO_2]_m$, but substitutions and/or vacancies are commonly found in the RS [4, 11, 12]. The relative periodicity $m$ between RS and triangular layers is usually incommensurate. It increases with the size of $Ae$, from $m=1.67$ for Ca, 1.82 for Sr and nearly 2 for Ba [13]. The RS transfer $x$ electrons to the $CoO_2$ slabs, i.e. there are $(1-x)$ holes in the Co $t_{2g}$ band (this definition of $x$ is equivalent to that in $Na_xCoO_2$). The value of $x$ cannot be exactly anticipated because of the non-stoichiometries, but, all $Ae$ being divalent, one can expect $x$ to increase as $1/m$ from Ba to Sr to Ca. Our single crystals were prepared by a standard flux method [11] and characterized by magnetic and transport measurements [14]. In BiBaCoO, there is a weak CW contribution to the susceptibility and $d\rho/dT$ is positive down to low temperatures [4]. For Sr and Ca, the CW contribution increases and there is an upturn in resistivity below 50K for Sr [5] and 200K for Ca [6,14]. Previous ARPES investigations reported spectra with small weight at the Fermi level and no angular dependence in (Bi,Pb)SrCoO [15,16] and a QP peak in (Bi,Pb)BaCoO [17]. We completely mapped the FS of BiBaCoO and establish that its electronic structure is very close to that of Na cobaltates. In BiSrCoO and BiCaCoO, the spectral weight at the Fermi level is gradually transferred to dispersive structures at high binding energies. This evolution, never observed in Na cobaltates, allows to clarify the lineshape structure.

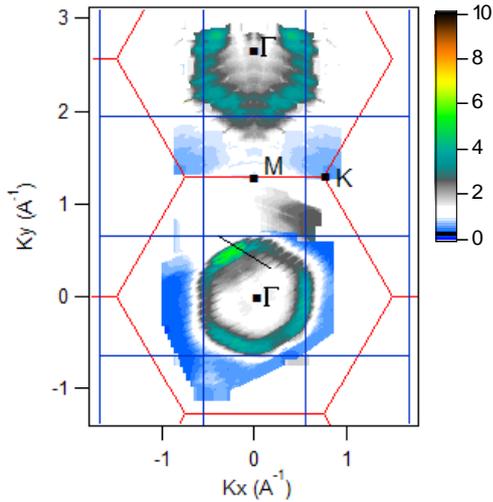

Fig. 1 : Fermi Surface of BiBaCoO measured at 15K with a photon energy of 100eV, obtained by integration of the spectral weight in a 15meV window around $E_f$. Thick red and thin blue lines indicate the BZ limits for triangular and RS planes. Data have been symmetrized with respect to $k_y = -\sqrt{3} \cdot k_x$ in the 1st BZ and $k_x=0$ in the 2nd BZ.

ARPES measurements were carried out at the SIS beamline of the Swiss Light Source (for BiBaCoO and BiSrCoO) and at the BACH beamline of ELETTRA (BiCaCoO). The FS of BiBaCoO is presented in Fig.1. The graph displays two hexagons, centered at (0,0) and (0, 2.55Å$^{-1}$). Thick red and thin blue lines indicate the limits of the Brillouin Zone (BZ) for the triangular and RS layers, respectively. The FS periodicity matches that of the triangular plane and the hexagonal shape of the FS also evidences its connection with the triangular lattice. This establishes that the $CoO_2$ are metallic and that, consequently, the comparison with Na cobaltates is meaningful. In Na cobaltates, a very similar hexagonal FS is indeed observed [18,19], where the hexagon corresponds to crossing of the singly degenerate $a_{1g}$ orbital of the $t_{2g}$ Co manifold [20]. As in Na cobaltates, we do not clearly observe other bands approaching the Fermi level, although LDA calculations [20] predict $e'_g$ bands to form small pockets in the ΓK direction.

The FS area is simply related by the Luttinger theorem to the number of holes per Co in the $a_{1g}$ band [18]. This gives us a way to estimate the actual hole doping of the Co layers, a crucial information in misfit cobaltates, where the doping cannot be simply deduced from the chemical composition. From measurements in three different samples, we estimate the Fermi wave vector along ΓM to be $k_f^{(M)} = 0.57 \pm 0.05$ Å$^{-1}$. Assuming an hexagonal FS, one expects $x = 1 - 2(k_f^{(M)}/ΓK)^2$, which leads to $x=0.7\pm0.05$. This value is consistent with the known properties of BiBaCoO that resemble those of $Na_xCoO_2$ with $x≈0.6$-$0.7$. However, it is *smaller* than the values reported for $Na_xCoO_2$ near $x=0.7$ ($k_f^{(M)} > 0.65$Å$^{-1}$ [18,19]), which suggests that a detailed comparison might reveal subtle differences in the properties of $CoO_2$ slabs as a function of their 3D environment.

We consider in Fig. 2 the electronic structure on a broader energy window. Two dispersing features are resolved, the narrow $a_{1g}$ QP peak near $E_f$ that traces the Fermi Surface (open triangle) and a broad shoulder at higher binding energies (filled triangle). They appear on top of a broad peak, centered at -0.8eV, which corresponds to the energy scale of the Co $t_{2g}$ manifold ($a_{1g} + 2\ e'_g$ bands) in the LDA calculation [20] and which is found with similar width and position in all cobaltates. The direction studied in Fig. 2 is not a high symmetry one (it is 25° off ΓK, as indicated in Fig.1), but we observe a similar structure in all directions when approaching the FS [21]. The universality of this feature is also established by the similar data reported in Na cobaltates, both for ΓM and ΓK directions [9,18].

In Fig. 2b and 2c, we present a complete analysis of these two features. Near the Fermi level, the spectra at constant energy (Momentum Distribution Curves, MDC) can be fitted by a lorentzian, yielding the QP dispersion indicated by red circles in Fig. 2b. As the QP peak vanishes, the MDC dispersion follows a nearly horizontal line (smaller size symbols), which is not meaningful and should not be confused with a "kink" in the dispersion. The QP itself can still be distinguished in the Energy Distribution Curve (EDC) spectra (open blue triangles in Fig. 2a and 2b) and seems to form a very narrow band of about 200meV. The slope near the Fermi level for this QP dispersion is $V_{QP} = 0.35\pm0.05$eV.Å [22], a similar order of magnitude than in Na cobaltates. This is a very small velocity (corresponding to $m^*=13 m_e$), but the band mass is itself rather large [20], implying $m^* \approx 3 m_{band}$. The MDC fit tracks again correctly the dispersion of the broad shoulder for E<-0.2eV. It is well aligned with the dispersion extracted from the maximum of the EDC spectra (blue triangles), which gives confidence in this analysis. Interestingly, the slopes of both dispersion extrapolate to the same $k_f$ value and the dispersion of the shoulder, $V_{sh}=1.3$eV.Å [22], is close to the bare band value for $a_{1g}$ [20]. We conclude that both the narrow peak and the broad shoulder belong to $a_{1g}$. Fig. 2c shows that the linewidth changes significantly in the two parts, increasing discontinuously from 30meV near $E_f$ to about 400meV in the shoulder. For a QP, the linewidth is inversely proportional to its lifetime. We note that the large width of the shoulder is difficultly compatible with this notion, as it would correspond to unphysically short lifetimes.

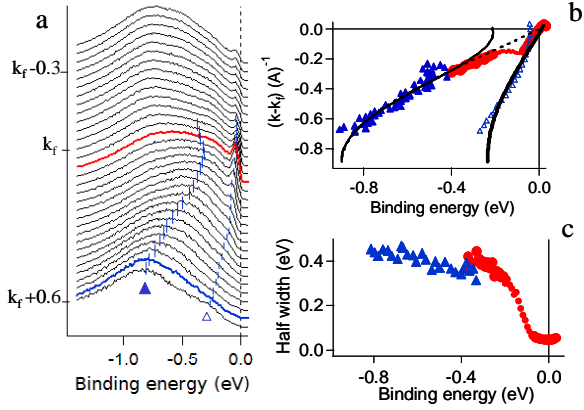

Fig. 2 : Dispersion in BiBaCoO along the direction shown in Fig.1. a) EDC spectra, spaced by 0.035Å$^{-1}$, with bars indicating the position of the shoulder (filled triangle) and QP peak (open triangle). b) Dispersion obtained from MDC fit (red circles) and EDC local maximum (blue triangles). Black lines are cosine fits of these dispersion. c) Half width at half maximum from MDC (width is normalized by slope of dispersion) and EDC spectra.

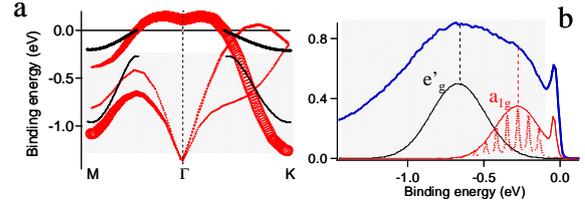

Fig. 3 : a) Red circles : sketch of the band structure (from ref. [23]). The size of the marker is proportional to the $a_{1g}$ character. Black lines : sketch for the dispersion of the two components identified in Fig. 2. Shaded area corresponds to incoherent excitations. b) Decomposition of a BiBaCoO spectrum near $E_f$ into $e'_g$ and $a_{1g}$ contribution. The $a_{1g}$ lineshape is sketched with a peak-dip-hump structure (see text).

Within a band structure approach, a "break" in the dispersion could be due to the opening of a gap at the crossing between two bands. The similar "2 peaks" structure reported for $a_{1g}$ in Na cobaltates was indeed recently interpreted as resulting from a large hybridization gap between $a_{1g}$ and $e'_g$ [9]. The band structure calculated by LDA is recalled in Fig. 3a [23], together with the dispersion of the two components extracted from Fig. 2b by a fit to cosine functions (we neglect here the small changes as a function of directions of the reciprocal space [9,21]). Calculations predict the opening of a large hybridization gap between $a_{1g}$ and $e'_g$ along ΓM [20,23], but not along ΓK, where it is forbidden by symmetry [20]. As the lineshape is essentially the same in the two directions, it is not likely that the hybridization gap is the origin of the structure. More fundamentally, this interpretation would not explain the change of dispersion and width in the two parts. These features correspond on the contrary very well to a "peak-dip-hump" (PDH) structure of the lineshape [10]. Fig. 3b sketches the leading QP peak and its incoherent tail, or "hump", represented as the envelop of multiple peaks corresponding, in a molecular picture, to different excited states. The hump is typically much broader than the QP and its width is not related to a lifetime but to the structure of the incoherent excitations. The size of the QP peak relatively to the hump is a measure of the strength of the interactions. The QP peak follows by definition the band dispersion renormalized by the interactions, while the hump may follow the bare band dispersion. These characteristics of dispersion and width perfectly fit our observations for the $a_{1g}$ structure. Recently, very similar "high-energy anomalies" were reported in many strongly correlated systems [24-26]. This structure therefore appears to be recurrent in correlated systems, although somewhat masked in cobaltates, because of the overlap with $e'_g$.

Indeed, there is probably a contribution of the two $e'_g$ bands in the low energy part of the spectrum of Fig. 3b. Its dispersion is not clearly observed in these experimental conditions [9]. We briefly note that this new interpretation of the lineshape structure could have important consequences for the search of $e'_g$ pockets in ARPES spectra. This debate received a lot of attention, because these pockets may play an essential role in some properties of cobaltates, like superconductivity [27]. Ref. [18] concluded that the pockets were pushed by correlations below $E_f$, to form a step at –0.2eV. This assignment is not as simple, if there is a PDH structure of the lineshape. The –0.2eV step could also be the bottom of the QP band, or possibly the top of a $e'_g$ "hump" with a vanishingly small QP peak.

The central question raised by this study is *the origin of the incoherent structure*. Fig. 4 reveals an evolution of the spectra through the Bi misfit family, which suggests answers. Fig. 4a shows that the QP peak is strongly suppressed in BiSrCoO and disappears in BiCaCoO. The qualitative trend of this evolution is maintained from 10 to 100K. A similar trend was recently reported in Pb-doped Bi misfit phases [28]. We further observe a dispersive "hump", very similar to that observed in BiBaCoO, as shown for BiCaCoO in Fig. 4b. Spectra near the top and the bottom of the band are reported in red (thick line) and blue (thin line) on Fig. 4a to indicate that this dispersion occurs on a similar energy scale in the three compounds. Therefore, the nearly insulating nature of BiCaCoO [6] does not result from a *shift* of the bands below $E_f$ but from a *transfer of spectral weight* from the QP peak to the hump. This behavior strongly supports the PDH interpretation, because the loss of weight at $E_f$ would be hardly understandable otherwise [15,28].

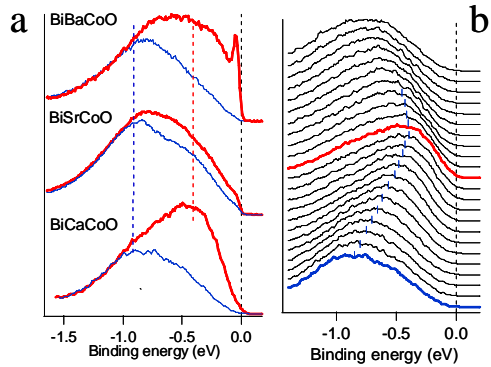

Fig. 4 : a) Spectra at 20K with minimum (thin blue line) and maximum (thick red line) weight at $E_f$ in BiBaCoO (see Fig.2a), BiSrCoO and BiCaCoO (see Fig.4b). b) Dispersion of the hump in BiCaCoO. Spectra are spaced by 0.063 Å$^{-1}$.

In misfits with Ba, Sr or Ca, one expects the doping to change because of the different relative periodicities $m$ [13]. Using the value $x=0.7$, determined by the present study for Ba, we may estimate the effective doping for the other compounds by using the ratio of periodicities. This yields $x=0.8$ for Sr and $x=0.9$ for Ca. We then interpret the evolution in Fig. 4 as the one towards the band insulator expected at $x=1$. In this limit of small hole-doping, correlation effects should become negligible and the suppression of the QP peak is very puzzling. The paradox is similar as that of the larger magnetic correlations near $x=1$, stressed in the introduction. In BiSrCoO and BiCaCoO, the magnetoresistance becomes large and negative [5], which also implies a strong coupling between carriers and magnetic excitations. We therefore conclude that elementary excitations have a magnetic character and must be a key ingredient for the PDH structure. One possibility would be the "spin-orbital polarons" proposed by Daghofer et al. [29] for Na$_x$CoO$_2$ near $x=1$. In this image, a hole (i.e. a Co$^{4+}$ ion) distorts the surrounding Co$^{3+}$ sites, allowing their excitation to high-spin states. This would create complex magnetic objects, with a spectral function markedly different from that of a simple QP. Dealing with elementary excitations extending over more than one site is a possible way to solve the puzzle of the large correlations in the small hole limit. Indeed, such "objects" would be stabilized, when they do not overlap too strongly.

To summarize, we have shown that the underlying electronic structure of misfit cobaltates is that of CoO$_2$ slabs. In BiBaCoO, a hexagonal FS is observed, like in Na cobaltates, the Fermi velocity is of the same order of magnitude and the lineshape structure is identical. The FS area corresponds to $x=0.7$ according to the Luttinger theorem. Increasing $x$ by replacing Ba with Sr or Ca leads to a strong suppression of the QP peak, never observed so far in Na cobaltates. We attribute this to a transfer of spectral weight towards an incoherent structure, which we identify as a broad hump dispersing at the bare band value, already present at lower $x$ values. This assignment is a step towards the understanding of correlation effects in cobaltates. It favors a picture, where elementary excitations are complex objects, formed by holes strongly coupled to magnetic and/or lattice excitations, and which possibly extend over more than one site. The lower conductivity of misfit phases with respect to Na phases, probably due to their larger disorder, likely explains that these objects tend to localize at high $x$ and become more easily detectable.

We thank A.F. Santander-Syro for his contribution during the initial stage of this work, H. Alloul, J. Bobroff, A. Georges, P. Mendels, F. Parmigiani, M. Rozenberg and M. Shi for useful discussions.